\documentclass[acmsmall]{acmart}
\usepackage{csquotes} %package to do inline or display quotes
\usepackage[normalem]{ulem} % for sout (strikethrough)
\AtBeginDocument{%
  \providecommand\BibTeX{{%
    \normalfont B\kern-0.5em{\scshape i\kern-0.25em b}\kern-0.8em\TeX}}}

\setcopyright{acmlicensed}
\acmJournal{PACMHCI}
\acmYear{2022} \acmVolume{6} \acmNumber{CSCW2} \acmArticle{340} \acmMonth{11} \acmPrice{15.00}\acmDOI{10.1145/3555760}

\received{April 2021}
\received[revised]{November 2021}
\received[accepted]{March 2022}


\usepackage{color}

\newcommand{\highlight}[1]{{#1}}

\usepackage{dirtytalk}
\begin{document}

%%
%% The "title" command has an optional parameter,
%% allowing the author to define a "short title" to be used in page headers.
\title[Understanding ML Practitioners' Data Documentation Perceptions, Needs, Challenges, and Desiderata]{Understanding Machine Learning Practitioners' Data Documentation Perceptions, Needs, Challenges, and Desiderata}

%%
%% The "author" command and its associated commands are used to define
%% the authors and their affiliations.
%% Of note is the shared affiliation of the first two authors, and the
%% "authornote" and "authornotemark" commands
%% used to denote shared contribution to the research.

\author{Amy K. Heger}
\email{v-amyheger@microsoft.com}
\affiliation{
 \institution{Microsoft}
\city{St. Louis, MO}
\country{USA}
}
\author{Liz B. Marquis}
\email{ebmorris@umich.edu}
\affiliation{
\institution{University of Michigan, School of Information}
\city{Ann Arbor, MI}
\country{USA}
}
\author{Mihaela Vorvoreanu}
\email{Mihaela.Vorvoreanu@microsoft.com}
\affiliation{
 \institution{Microsoft}
\city{Redmond, WA}
\country{USA}
}
\author{Hanna Wallach}
\email{wallach@microsoft.com}
\affiliation{
 \institution{Microsoft}
\city{New York, NY}
\country{USA}
}
\author{Jennifer Wortman Vaughan}
\email{jenn@microsoft.com}
\affiliation{
 \institution{Microsoft}
\city{New York, NY}
\country{USA}
}

% \author{Lars Th{\o}rv{\"a}ld}
% \affiliation{%
%   \institution{The Th{\o}rv{\"a}ld Group}
%   \streetaddress{1 Th{\o}rv{\"a}ld Circle}
%   \city{Hekla}
%   \country{Iceland}}
% \email{larst@affiliation.org}

%%
%% By default, the full list of authors will be used in the page
%% headers. Often, this list is too long, and will overlap
%% other information printed in the page headers. This command allows
%% the author to define a more concise list
%% of authors' names for this purpose.
\renewcommand{\shortauthors}{Amy Heger et al.}

%%
%% The abstract is a short summary of the work to be presented in the
%% article.
\begin{abstract}
  Data is central to the development and evaluation of machine learning (ML) models. However, the use of problematic or inappropriate datasets can result in harms when the resulting models are deployed. To encourage responsible AI practice through more deliberate reflection on datasets and transparency around the processes by which they are created, researchers and practitioners have begun to advocate for increased data documentation and have proposed several data documentation frameworks. However, there is little research on whether these data documentation frameworks meet the needs of ML practitioners, who both create and consume datasets. To address this gap, we set out to understand ML practitioners' data documentation perceptions, needs, challenges, and desiderata, with the ultimate goal of deriving design requirements that can inform future data documentation frameworks. We conducted a series of semi-structured interviews with 14 ML practitioners at a single large, international technology company. We had them answer a list of questions taken from datasheets for datasets~\cite{gebru2018datasheets}. Our findings show that current approaches to data documentation are largely ad hoc and myopic in nature. Participants expressed needs for data documentation frameworks to be adaptable to their contexts, integrated into their existing tools and workflows, and automated wherever possible. Despite the fact that data documentation frameworks are often motivated from the perspective of responsible AI, participants did not make the connection between the questions that they were asked to answer and their responsible AI implications. In addition, participants often had difficulties prioritizing the needs of dataset consumers and providing information that someone unfamiliar with their datasets might need to know. Based on these findings, we derive seven design requirements for future data documentation frameworks such as more actionable guidance on how the characteristics of datasets might result in harms and how these harms might be mitigated, more explicit prompts for reflection, automated adaptation to different contexts, and integration into ML practitioners' existing tools and workflows.
\end{abstract}

%%
%% Keywords. The author(s) should pick words that accurately describe
%% the work being presented. Separate the keywords with commas.

%%
%% This command processes the author and affiliation and title
%% information and builds the first part of the formatted document.
\maketitle

\section{Introduction}

Data plays a pivotal role in the development and evaluation of the machine learning (ML) models that are now pervasive in criminal justice~\cite{chouldechova2017fair}, public policy~\cite{veale2018fairness, hacker2018teaching}, policing~\cite{selbst2017disparate, haque2019exploring}, healthcare~\cite{lee2021included}, and other critical domains. However, a model is only as good as the datasets with which it is developed and evaluated, and datasets can be problematic or inappropriate in different ways~\cite{PRB+20}. For example, a lack of appropriate representation of different groups of people can lead to models that exhibit performance disparities. Prominent datasets used to develop and evaluate facial recognition systems have been shown to lack appropriate representation of people with darker skin tones, causing commercially available facial analysis systems to perform poorly for darker-skinned women~\cite{BG18}. Similarly, common object recognition datasets have been shown to lack appropriate geographical representation, leading to poor performance of object recognition systems on household items more common in countries with lower incomes~\cite{VMWM19}. As another example, spurious correlations and other unanticipated anomalies in training datasets can result in models that fail to generalize~\cite{GJM+20}. Researchers found that a model trained to detect the presence of pneumonia from X-ray scans picked up on hospital-specific metal tokens present on each scan~\cite{ZBL+18}. Since the prevalence of pneumonia varied widely for different hospitals in the model's training dataset, these tokens were a good indicator of whether or not pneumonia was present, but led to poor performance for scans from hospitals that were not represented in the training dataset. In addition, subjectivity in dataset labels and inaccurate notions of ground truth can result in models with misleading outputs~\cite{AW15,HFG19}. Furthermore, all of these problems are exacerbated by the common practice of sharing and using datasets with little attention to their contents and characteristics or to the processes that led to their creation~\cite{HSH+21}. \looseness=-1

To encourage more deliberate reflection on datasets and transparency around the processes by which they are created, researchers and practitioners have begun to advocate for increased data documentation~\cite{gebru2018datasheets,holland2018dataset,bender2018data,aboutml19,HSH+21}. Proposed data documentation frameworks include datasheets for datasets~\cite{gebru2018datasheets}, dataset nutrition labels~\cite{holland2018dataset}, and data statements for natural language processing datasets~\cite{bender2018data}. Data documentation has the potential to help dataset creators think through the underlying assumptions, potential risks, and implications of use of their datasets, encouraging increased reflection and accountability. Simultaneously, data documentation can help dataset consumers (i.e., those using a dataset to develop or evaluate their models) make informed decisions about whether specific datasets meet their needs, effectively acting as a mode of communication and collaboration between dataset creators and dataset consumers~\cite{gebru2018datasheets, miceli2021documenting}. For these reasons, it has been argued that data documentation is crucial for responsible AI practice---that is, the development, evaluation, and deployment of AI systems (including ML models) in ways that prioritize principles like transparency, fairness, safety, reliability, and privacy~\cite{rakova2021responsible,jobin2019global}.

Despite the promise of data documentation frameworks, little research has been done on whether they meet the needs of ML practitioners, particularly in industry where it is often difficult to translate general responsible AI principles into concrete practices~\cite{madaio2020co, holstein2019improving, rakova2021responsible, veale2018fairness}. Existing data documentation frameworks do not distinguish between industrial and academic use cases, although an industry practitioner creating an internal proprietary dataset may very well have different needs compared, for example, to an academic researcher creating a public benchmarking dataset. Industry practitioners have existing workflows and tools, and previous research has illustrated the importance of ensuring that responsible AI efforts are aligned with these tools and workflows~\cite{madaio2020co}.  Furthermore, datasets are often created, maintained, and used collaboratively by many stakeholders such as product managers, data scientists, and software engineers, possibly working on different teams or in different organizations, all of whom interact with datasets at different points in time~\cite{holstein2019improving, passi2018trust, zhang2020data}. Data documentation is therefore not only a collaboration artifact; the processes by which it is created are also collaborative~\cite{miceli2021documenting, aboutml19, neang2021data}.\looseness=-1

Our ultimate goal is to derive design requirements that can inform future data documentation frameworks. We do so by understanding what ML practitioners' experiences are like with a hands-on data documentation exercise. Two research questions drive our study:

\begin{itemize}
    \item \textbf{RQ1} How do industry ML practitioners currently approach data documentation?
    \item \textbf{RQ2} What are industry ML practitioners' data documentation perceptions, needs, challenges, and desiderata?\looseness=-1
\end{itemize}

Our first research question enables us to understand ML practitioners' contexts and current approaches to data documentation, if any. This provides necessary background for our second research question, which is the primary focus of this paper.

We conducted a series of semi-structured interviews with 14 ML practitioners at a single large, international technology company in roles such as ML scientist, data scientist, project manager, software engineer, and researcher. In preliminary interviews, we asked participants to describe their current approaches to data documentation and any challenges that they had encountered (RQ1). Next, we gave participants the full list of questions from datasheets for datasets~\cite{gebru2018datasheets} and asked them to answer the questions to the best of their abilities with respect to a dataset with which they had worked recently. We then conducted follow-up interviews to understand their data documentation perceptions, needs, challenges, and desiderata (RQ2).\looseness=-1

Previous research has looked at data documentation from the perspective of dataset consumers~\cite{B21}. Our study is complementary in that it focuses on dataset creators. Specifically, our focus is on improving data documentation as a collaboration artifact (i.e., a boundary object~\cite{lee2007boundary}). By improving data documentation as a collaboration artifact, we contribute to improved communication and collaboration between dataset creators and dataset consumers, and to the larger goal of responsible AI. However, as mentioned above, the processes by which data documentation is created are also collaborative~\cite{miceli2021documenting, aboutml19, neang2021data}. We therefore derive design requirements that relate to these collaborative processes, as well as design requirements that relate to data documentation as a collaboration artifact.\looseness=-1

Our study should not be viewed as an evaluation of datasheets for datasets~\cite{gebru2018datasheets}. We used the datasheets for datasets questions only as an exercise to support reflection and the elicitation of participants' data documentation perceptions, needs, challenges, and desiderata. We wanted to ensure that participants would be sufficiently well informed about what creating data documentation might entail, and reasoned that a hands-on exercise would accomplish this goal more effectively than simply reading through a typical data documentation framework. We note that the way in which participants answered the questions does not reflect the way that datasheets for datasets should or would be used in practice.  For example, while \citet{gebru2018datasheets} note that the list of questions is not meant to be prescriptive and should be adapted to different contexts, we provided the full list to participants as a way to understand which questions they found to be most relevant. In addition, we omitted guidance on when each question should be answered and provided the questions in a single Word document rather than integrating them into participants' existing tools and workflows. As discussed below, we believe that many of our findings, as well as the design requirements that we derived, are applicable to other data documentation frameworks since most of these frameworks share common attributes.\looseness=-1

Overall, participants perceived data documentation as useful, and listed many benefits, including facilitating collaboration via dataset discovery and helping with onboarding onto a dataset. However, participants took ad-hoc and myopic approaches to data documentation, prioritizing information about whether they or others \emph{could} use a dataset for a specific purpose---that is, whether it was approved for use---over a deeper consideration of whether they or others \emph{should} use it and what might go wrong if they were to do so.
Strikingly, although data documentation frameworks are often motivated from the perspective of responsible AI, participants did not make the connection between the questions that they were asked to answer and their responsible AI implications. This suggests the need to provide more actionable guidance on how the characteristics of datasets might result in harms and how these harms might be mitigated, as well as more explicit prompts for reflection.
Participants also faced other challenges while documenting their chosen datasets, including difficulties answering some of the questions. They were unsure about the right level of granularity with which to answer the questions, typically did not seek assistance from others when they did not know how to answer the questions, and in some cases---especially with streaming datasets that consist of data collected continuously over time---were uncertain about what counts as a dataset. Further complications arose due to differences across participants' contexts. Our findings corroborate a point made by \citet{gebru2018datasheets}: there is no one-size-fits-all approach to data documentation. We therefore discuss the need for data documentation frameworks to be adaptable---perhaps automatically---based on context-specific factors such as whether a dataset is static or streaming or how broadly a dataset will be distributed. In terms of desiderata, participants wanted data documentation frameworks to have interactive capabilities, integration into their existing tools and workflows, and automation to keep data documentation up to date. Based on these findings, we derive seven design requirements that can inform future data documentation frameworks. \looseness=-1

While not the primary focus of this paper, we are continuing to build on these results to inform data documentation practices in industry, including generating an updated data documentation template, revised to address participants' needs and better align with our derived design requirements. The most up-to-date information can be found on our project website.\footnote{\url{http://aka.ms/datadoc}}

\section{Related work}  \label{Literature}

As mentioned above, researchers and practitioners have begun to advocate for increased data documentation and have proposed several data documentation frameworks~\cite{gebru2018datasheets,holland2018dataset,bender2018data,aboutml19,HSH+21}. While these data documentation frameworks vary in terms of their target audience and whether they are domain specific or general purpose, they share the goal of encouraging more deliberate reflection on datasets and transparency around the processes by which they are created.

Data statements~\cite{bender2018data} were proposed as a data documentation framework for natural language processing datasets. A data statement includes context to allow consumers of a dataset to understand how experimental results obtained using it might generalize, considerations for deploying models developed or evaluated using it, and any societal biases that might be reflected in it. The content is tailored to language-based data and includes information such as speaker demographics and language variety.\looseness=-1

In contrast to data statements' focus on natural language processing datasets, dataset nutrition labels~\cite{holland2018dataset, Chmielinski2020DNL2} and datasheets for datasets~\cite{gebru2018datasheets} are two more general data documentation frameworks. Dataset nutrition labels~\cite{holland2018dataset,Chmielinski2020DNL2} were inspired by nutrition labels on food. They were designed to highlight the ``ingredients'' of a dataset, covering metadata and provenance, as well as information about any unique or anomalous characteristics. The second-generation dataset nutrition label tool is specifically designed to support data scientists searching for datasets to use when developing or evaluating their models.  As such, a dataset nutrition label emphasizes information about a dataset's intended use cases, including alerts that highlight licensing restrictions and the potential for harms. \looseness=-1

Developed contemporaneously, datasheets for datasets~\cite{gebru2018datasheets} were inspired by standard documentation practices in the electronics industry. They were designed to encourage dataset creators to reflect on the choices made throughout the dataset lifecycle, and to help dataset consumers---including data scientists searching for datasets to use when developing or evaluating their models---make more informed decisions.  Datasheets for datasets consist of questions that dataset creators should answer when documenting their datasets.  The questions are grouped into sections organized around seven stages of the dataset lifecycle: motivation, composition, collection, processing/cleaning/labeling, use, distribution, and maintenance. The authors note that these questions are not meant to be prescriptive, but should be adapted to different contexts and integrated into existing tools and workflows as appropriate. The questions touch upon general characteristics (``Is there a label or target associated with each instance?'') and limitations (``Is any information missing from individual instances?''), as well as factors related to responsible AI like fairness (``Describe how these [demographic] groups are identified and provide a description of their respective distributions within the dataset''), privacy  (``Does the dataset contain data that might be considered confidential?''), potential legal implications (``Did the individuals in question consent to the collection and use of their data?''), and ethics (``Who was involved in the data collection process [e.g., students, crowdworkers, contractors] and how were they compensated [e.g., how much were crowdworkers paid]?''). Example questions are provided in Table~\ref{table:questions}. Since datasheets for datasets were designed to prompt reflection, answers are intentionally not automated.\looseness=-1

We used the full list of questions from datasheets for datasets in our study as an exercise to support reflection and the elicitation of participants' data documentation perceptions, needs, challenges, and desiderata. These questions are an appropriate choice for our study because they are generally applicable across domains and, unlike dataset nutrition labels, are not tied to a specific interface or tool. We believe that many of our findings, as well as the design requirements that we derived, are applicable to other data documentation frameworks since most of these frameworks share common attributes. For example, like datasheets for datasets, data statements and dataset nutrition labels involve manually collecting information from the creators of a dataset about its motivation, the processes by which it was created, and factors related to responsible AI such as the demographics of the people whose data is included in it. In fact, some of the questions included in the second-generation dataset nutrition label tool were taken directly from datasheets for datasets~\cite{Chmielinski2020DNL2}.\looseness=-1

Other researchers have taken different perspectives on data documentation. \citet{miceli2021documenting} emphasize the need for a reflexive approach and the importance of accounting for the conditions under which a dataset was created.  In the context of computer vision datasets, they advocate for exposing the implications of asymmetrical power relationships between those who request or commission datasets and those who label them, questioning the assumptions behind labels such as race and gender, and acknowledging that ``ground truth'' is often subjective.
\citet{HSH+21} view datasets as a form of infrastructure and propose incorporating practices from software engineering into the documentation, oversight, and maintenance of datasets.  They suggest breaking data documentation down into five components that mirror the forms of documentation typically produced as part of the software development lifecycle such as dataset requirements specifications and dataset design documents.\looseness=-1

Moving beyond data documentation specifically, researchers and practitioners have proposed frameworks analogous to those described above for documenting ML models and AI systems~\cite{mitchell2019model,arnold2019factsheets}. While these documentation frameworks are not focused on datasets alone, they include questions about the datasets that are used for model development and evaluation. For example, the authors of FactSheets~\cite{arnold2019factsheets} advocate for linking to datasheets for such datasets.

Regardless of the specific framework, data documentation is inherently collaborative. Much like the processes by which datasets are created, the processes by which data documentation is created require communication and collaboration between many stakeholders, all of whom interact with datasets at different points in time~\cite{passi2018trust,miceli2020between,miceli2021documenting}.  For example, different stakeholders might be involved in creating the initial specification for a dataset, collecting data instances, labeling those instances, processing and cleaning the dataset, and maintaining the dataset~\cite{HSH+21}.
In addition, data documentation can facilitate communication and collaboration between dataset creators, dataset consumers using datasets to develop or evaluate their models, and potentially even the people who may be affected by those models when deployed~\cite{aboutml19,gebru2018datasheets}.
In this way, data documentation can serve as a boundary object~\cite{star1989institutional} that facilitates communication and collaboration between the stakeholders who interact with a dataset throughout its lifecycle and even beyond, as systems based on that dataset are deployed. Data documentation frameworks and other conceptual schemes used for indexing and classifying information can help mitigate the complex demands of articulation work~\cite{schmidt2008taking}---that is, the work required to coordinate and manage the stakeholders and tasks involved in distributed cooperative work~\cite{strauss1988articulation}. Therefore, it is important for the CSCW community to understand how data documentation is created in order to inform future data documentation frameworks and their adoption.\looseness=-1

Data documentation frameworks are part of a larger push toward responsible AI. While there has been a proliferation of responsible AI principles proposed in recent years~\cite{jobin2019global}, translating these general principles into concrete practices is often nontrivial~\cite{vitak2016beyond, mittelstadt2019ai,krafft2020defining,madaio2020co,holstein2019improving, rakova2021responsible}. This is particularly true in industry, where ML practitioners are seldom given the resources or incentives to do so and face other organizational and cultural barriers~\cite{madaio2020co,rakova2021responsible, holstein2019improving, mittelstadt2019ai}.

Currently, little is known about whether data documentation frameworks meet the needs of ML practitioners. While the CHI and CSCW communities have investigated the collaborative work of data science and ML more broadly~\cite[e.g.,][]{Mao2019datascicollab,holstein2019improving,hou2017NPO,muller2019dsworkers,passi2018trust, boyd2020ethical} and proposed systems to support collaboration within data science teams \cite[e.g.,][]{crowston2019midst, katsis2019modellens}, data documentation frameworks have not been explored in depth. One notable exception is the recent work of Boyd~\cite{B21}, who explored how datasheets for datasets can enable dataset consumers to identify ethical issues with datasets. In a think-aloud study, in which data scientists and ML engineers were presented with a dataset and a hypothetical scenario for its use, Boyd found that participants who were given a datasheet for that dataset (created by the author of the study) mentioned ethical issues earlier and more often than participants who were not. Our research complements this work, focusing on the perspective of dataset creators, since the benefits of data documentation necessarily depend on its quality. A survey focused on data scientists' collaboration practices~\cite{zhang2020data}, which included a section on data documentation, found gaps in their data documentation approaches. Even though about 60\% of participants, on average, expected their datasets to be reused, about a third to a quarter did not do ``even something basic like adding column labels to datasets''~\cite[][page 15]{zhang2020data}. To address such gaps, we need a comprehensive understanding of ML practitioners' data documentation perceptions, needs, challenges, and desiderata. This paper contributes to that goal.

\section{Methods}    \label{Method}

To answer our research questions and derive design requirements that can inform future data documentation frameworks, we took a three-step approach. First, we conducted preliminary interviews with 14 ML practitioners to understand their contexts and current approaches to data documentation, if any (RQ1). We then asked them to complete a data documentation exercise in which they were asked to fill out a form containing the full list of questions from datasheets for datasets~\cite{gebru2018datasheets}. Finally, we conducted follow-up interviews to understand their data documentation perceptions, needs, challenges, and desiderata, as informed by the data documentation exercise (RQ2). All interviews were conducted virtually on a video conferencing platform. Participation was voluntary and participants did not receive compensation. The study was approved by our institution's IRB.\looseness=-1

\subsection{Preliminary Interviews}   \label{Prelim}

As described above, the preliminary interviews were intended to help us understand participants' contexts and current approaches to data documentation, if any. We used a semi-structured-interview approach to probe on the following topics: the types of datasets with which participants worked, the dataset lifecycle, and participants' processes for creating datasets; participants' current approaches to data documentation; benefits and drawbacks of data documentation in general and of participants' specific approaches to data documentation; participants' ideas for their ideal data documentation frameworks.\looseness=-1

\subsection{Data Documentation Exercise}   \label{Exercise}

Following the preliminary interviews, we emailed participants a form to fill out. The form was a 15-page Word document containing the full list of questions from datasheets for datasets~\cite{gebru2018datasheets}---that is, 51 unique questions grouped into sections organized around seven stages of the dataset lifecycle. See Table~\ref{table:questions} for example questions and see the supplemental material for the form provided to participants. We asked each participant to answer the questions with respect to a dataset with which they had worked recently. We additionally provided space for participants to add comments under each question, and to specify whether they currently document the information covered by each question and, if so, where. Participants were instructed to {``answer each question to the best of your ability''} and to {``use the comments box to document your thought process and provide feedback on each question's clarity, usefulness, effort required to answer, etc.''} Roughly one third of participants misunderstood or ignored the instructions regarding the comment boxes, even misidentifying them as part of the datasheets for datasets framework rather than a component of our research.\looseness=-1

\begin{table}[]
\begin{tabular} {p{2cm}p{11.3cm}}
\toprule
\textbf{Section} & \textbf{Example questions}
   \\ \midrule
Motivation \newline (3 questions) &
  1.1  For what purpose was the dataset created? Was there a specific task in mind? Was   there a specific gap that needed to be filled? Please provide a description. \newline
  1.2  Who created the dataset (e.g., which team, research group) and on   behalf of which entity (e.g., company, institution, organization)? \vspace{.15cm}
   \\
Composition \newline (16 questions) &
2.1  What do the instances that comprise the dataset represent (e.g., documents, photos, people, countries)? Are there multiple types of instances (e.g., movies, users, and   ratings; people and interactions between them; nodes and edges)? Please provide a description. \newline
2.7  Are relationships between individual instances made explicit (e.g., users' movie ratings, social network links)? If so, please describe how these relationships are made explicit.\newline
2.14  Does the dataset identify any demographic groups, groups that are protected by law, or groups that may be considered sensitive (e.g., age,   gender, disability status)? If so, please describe how these groups are identified and provide a description of their respective distributions within the dataset. \vspace{.15cm}
   \\
Collection Process \newline (11 questions) &
3.2  What mechanisms or procedures were used to collect the data (e.g., hardware apparatus or sensor, manual human curation, software program, software API)? How were these mechanisms or procedures validated? \newline
3.8  Were the individuals in question notified about the data collection? If so, please describe (or show with screenshots or other information) how notice was provided, and provide a link or other access point to, or otherwise reproduce, the exact language of the notification itself. \vspace{.15cm}
   \\
Preprocessing, \newline Cleaning, \newline \& Labeling \newline (3 questions) &
  4.1  Was any preprocessing/cleaning/labeling of the data done (e.g., discretization or bucketing, tokenization, part-of-speech tagging, SIFT feature extraction,  removal of instances, processing of missing values)? If so, please provide a description. If not, you may skip the remainder of the questions in this section. \vspace{.05cm}
   \\
Uses \newline (5 questions) &
5.3  What (other) tasks could the dataset be used for? \newline
 5.5  Are there tasks for which the dataset should not be used? If so, please provide a description. \vspace{.15cm}
  \\
Distribution \newline (6 questions) &
6.1  Will the dataset be distributed to third parties outside of the entity (e.g., company, institution, organization) on behalf of which the dataset was created? If so, please provide a description. \newline
6.4  Will the dataset be distributed under a copyright or other intellectual property (IP) license, and/or under applicable terms of use (ToU)? If so, please describe this license and/or ToU, and provide a link or other access point to, or otherwise reproduce, any relevant licensing terms or ToU, as well as any fees associated with these restrictions. \vspace{.15cm}
   \\
Maintenance \newline (7 questions) &
  7.2  How can the owner/curator/manager of the dataset be contacted (e.g., email address)? \newline
 7.4  Will the dataset be updated (e.g., to correct labeling errors, add new instances, delete instances)? If so, please describe how often, by whom, and how updates will be communicated to users (e.g., mailing list, GitHub)?
   \\ \bottomrule
\end{tabular}
\caption{Example datasheets for datasets questions~\cite{gebru2018datasheets} \label{table:questions}.}
\end{table}

Along with the list of questions, an example datasheet was provided to participants. We chose to use the Gender Inclusive Coreference (GICoref) dataset's datasheet with permission from the authors~\cite{cao2019toward}. We chose this datasheet because it was publicly available, detailed, and well executed.

The data documentation exercise was intended to support reflection and the elicitation of participants' data documentation perceptions, needs, challenges, and desiderata. We asked participants to complete this exercise so that they would be familiar with at least one data documentation framework and would therefore have an informed experience about which to speak in the follow-up interviews. Participants were asked to focus on datasets with which they had worked recently so that our findings would be better grounded in the realities of their day-to-day work. In our follow-up interviews, we referred to participants' answers to the datasheets for datasets questions as a ``datasheet,'' but we reiterate that our methods do not reflect the instructions, format, and adaptations described by~\citet{gebru2018datasheets}.\looseness=-1

\subsection{Follow-up Interviews}   \label{Followup}

The benefits of data documentation depend on the ability and willingness of dataset creators to produce high-quality data documentation. As a result, it is critical to understand their data documentation perceptions, needs, challenges, and desiderata (RQ2). If data documentation frameworks don't meet the needs of dataset creators, they may be unwilling to adopt them. After participants completed the data documentation exercise, they participated in semi-structured interviews, each lasting approximately 60 minutes. We asked specific questions about the processes that participants used when answering the datasheets for datasets questions (e.g., did you refer to other resources to find answers?), any challenges that they had encountered (e.g., what was difficult?), and their desiderata for future data documentation frameworks (e.g., what would you do differently?).

\subsection{Sampling Strategy}   \label{Sampling}

We used a criterion sampling strategy~\cite{maxwell2012qualitative}, with two main sampling criteria: 1) participants had to be ML practitioners who worked with datasets, and 2) participants had to work with static datasets. Unlike streaming datasets that consist of data collected continuously over time, static datasets are updated only rarely. Datasheets for datasets were designed for static datasets. Despite these criteria, multiple participants ended up choosing to focus on streaming datasets, as discussed in more detail in Section~\ref{ParticipantsData}. All participants were from a single large, international technology company that regularly integrates ML models into its products and services. We focused on recruiting participants in roles such as ML scientist, data scientist, project manager, software engineer, and researcher. Where possible, we tried to recruit participants who worked together, so that we could gather multiple perspectives on their teams' current approaches to data documentation. Once we had recruited a few participants through broad announcements, we used snowball sampling to recruit more.\looseness=-1

\subsection{Data Analysis} \label{AnalysisPlan}

We conducted a thematic analysis~\citep{braunclarke2006} of the interview transcripts and reviewed participants' answers to the datasheets for datasets questions. When reviewing participants' answers, we focused on identifying any challenges that they might have encountered when answering the questions, as well as any unmet needs. For example, we observed that participants answered the questions differently or encountered different challenges when answering the questions depending on their contexts (e.g., the types of datasets with which they worked, their scopes of dataset distribution). To surface these kinds of observations, we triangulated participants' answers to the datasheets for datasets questions with their subsequent reflections in the follow-up interviews.

Our thematic analysis followed an inductive and iterative process, guided by our research questions. One of the authors conducted the bulk of the analysis, in consultation with two of the other authors, who had conducted the interviews. The analysis began with repeated readings of the transcripts and the extraction of any codes that aligned with the concepts in our research questions. As the analysis progressed and new codes emerged, they were added to the codebook. Data were coded at the sentence or paragraph level, focusing on key ideas. Over the course of several meetings, all of the authors discussed the codes and organized them into themes and subthemes. For example, one major theme that emerged was ``desiderata,'' which relates to our second research question and captures participants' desires for future data documentation frameworks. Under this theme, the subtheme of ``integration'' was created by grouping together codes that covered having central locations for data documentation, tracking metadata, integrating data documentation into existing tools and workflows, and linking to scripts and other resources from data documentation. The organization of our findings in Section~\ref{Findings} reflects the structure of the main themes that we identified.\looseness=-1

Based on our findings, we derived seven design requirements that can inform future data documentation frameworks. These design requirements emerged from unmet needs, challenges, and confusion that came across in participants' answers to the datasheets for datasets questions, triangulated with their subsequent reflections in the follow-up interviews. They were also informed by participants' explicit desiderata for future data documentation frameworks, although there is not a one-to-one mapping between these desiderata and our design requirements. For example, satisfying participants' desire for automation would conflict with the goal of prompting reflection on datasets. When deriving our design requirements, we followed best practices from UX design~\cite{Nielsen2001,Gocza2015,Yeykelis2018}  that caution against the direct translation of participants' desiderata into design requirements.\looseness=-1

% TODO: If we want to mention that we additionally used the data to refine the list of questions, we could do that here.

\section{Findings} \label{Findings}

In this section, we present our findings, organized by the main themes that we identified via our thematic analysis. We begin with a description of the 14 participants and the data that we collected from them (Section~\ref{ParticipantsData}). Next, we discuss participants' current approaches to data documentation, answering our first research question (Section~\ref{RQ1}). Then, to address our second, and primary, research question, we describe participants' data documentation perceptions (Section~\ref{OverallPerceptions}), needs and challenges (Section~\ref{Challenges}), and desiderata (Section~\ref{Desiderata}).\looseness=-1

\subsection{Participants and Data} \label{ParticipantsData}

We recruited 15 participants on four teams at a single large, international technology company to participate in our preliminary interviews. One participant (P13) chose not to continue with the study, so we ultimately analyzed data from 14 participants. We reached data saturation, as evidenced by the recurrence of key ideas in participants' interviews such as the need for data documentation frameworks to be automated wherever possible, participants' failures to make the connection between the datasheets for datasets questions and their responsible AI implications, and participants' frustrations with the length of the data documentation exercise. Participants' roles and teams, as well as information about the datasets on which they chose to focus are presented in Table~\ref{table:participants}.\looseness=-1

\begin{table}[]
\begin{tabular}{lllllll}
  \toprule
\textbf{Team}   & \textbf{Participant} & \textbf{Role}              & \textbf{Type} & \textbf{Source} & \textbf{Distribution} & \textbf{Content}    \\
\midrule
Team A & P1          & ML Scientist      & Static    & Public      & Internal          & Text       \\
       & P2          & ML Scientist      & Streaming & Private     & Internal          & Text/Image \\
       & P3          & Data Scientist    & Streaming & Private     & Internal          & Text/Image \\
       & P4          & ML Scientist      & Streaming & Private     & Internal          & Event logs \\
       \midrule
Team B & P5          & Software Engineer & Static    & Public      & Internal          & Text       \\
       & P8          & Software Engineer & Streaming & Public      & Internal          & Text       \\
       & P11         & Software Engineer & Streaming & Private     & Internal          & Text       \\
       & P15         & Data Scientist    & All       & Both        & Internal          & Text       \\
       \midrule
Team C & P6          & Data Scientist    & Streaming & Private     & Internal          & Metrics    \\
       & P7          & Project Manager   & Streaming & Private     & Internal          & Metrics    \\
       & P12         & Software Engineer & Streaming & Private     & Internal          & Metrics    \\
       & P14         & Data Scientist    & Streaming & Private     & Internal          & Metrics    \\
       \midrule
Team D & P9          & Research Intern   & Static    & Public      & External          & Text/Audio \\
       & P10         & Researcher        & Static    & Public      & External          & Text/Audio\\
\bottomrule
\end{tabular}
\caption{Participants and the datasets on which they chose to focus for the data documentation exercise. \label{table:participants}}
\end{table}

We conducted 28 interviews (i.e., one preliminary interview and one follow-up interview with each of the 14 participants) over a period of two months during the summer of 2020, totaling 23 hours and nine minutes. On average, there were 14 days between each participant's preliminary and follow-up interviews (range 6--25 days). Each preliminary interview lasted 50 minutes on average (range 32--62 minutes) and each follow-up interview last 52 minutes on average (range 28--62 minutes).\looseness=-1

Participants completed 12 ``datasheets,''  and we included 11 of these in our data analysis. Two participants (P9, P10) chose to focus on the same dataset and answered the datasheets for datasets questions almost identically. We therefore analyzed only one copy of their answers as the other contained no unique information. These participants gave separate follow-up interviews that captured their distinct perspectives. In addition, two participants did not answer the datasheets for datasets questions as instructed. One participant (P8) did not complete the data documentation exercise between their preliminary and follow-up interviews, but provided feedback on the questions and gave potential answers during their follow-up interview. The other participant (P7) did not answer the datasheets for datasets questions for a specific dataset, but instead made a chart that specified whether each question \textit{``applies for most cases,'' ``applies in certain cases,''} or \textit{``does not apply,''} and listed different datasets and their reasoning in these columns. They also discussed the questions and gave potential answers during their follow-up interview.

All participants, except for
two researchers on Team D, chose to focus on datasets that were not distributed outside of the company, which participants viewed as the definition of external distribution. The two researchers, in contrast, were creating a dataset for the purpose of sharing it online as a resource for others to use when developing or evaluating ML models. Four participants chose to focus on static datasets, while the remaining 10 participants chose to focus on streaming datasets. This was not what we had expected, but in the absence of a reminder, it is possible that participants forgot our instructions and simply chose to focus on datasets with which they had worked recently, which, for 10 participants, were streaming datasets. Eight participants used private datasets sourced from the company's users and devices, while the remaining six participants used public datasets sourced from the web that had either already been compiled or that they had scraped themselves. The content of the datasets ranged from text snippets to aggregate business metrics.\looseness=-1

\subsection{ML Practitioners' Current Approaches to Data Documentation} \label{RQ1}

The themes in this section relate to our first research question---that is, how do industry ML practitioners currently approach data documentation?
We describe participants' current approaches at the team level. These findings provide necessary background for our second research question.\looseness=-1

\subsubsection{Ad-Hoc Approaches} \label{CurrentPractices}

Participants' current approaches to data documentation are largely characterized by a set of contradictions: information was documented in dedicated locations and in documents created for other purposes; participants saw their approaches as incipient and informal, yet sufficient for their own needs, while also saying that they should do more.\looseness=-1

The dedicated locations that participants reported using for data documentation included wikis, team notebooks, and text files stored in GitHub repositories. These locations were used to document all sorts of information, not only information about datasets. For example, Team C maintained a wiki to document the data streams that they used. The wiki included links to the streams themselves, as well as scripts and workflows; descriptions of data columns and pipeline dependencies; and team contact information. Multiple participants on this team shared that they referenced the wiki to find the information that they needed to answer the datasheets for datasets questions. Team A had a notebook that was supposed to be a shared resource for keeping track of datasets, but it was not consistently used or referenced. Team B created their own datasheet to document data provenance, which they stored as a text file in their GitHub repositories. This informal, team-initiated practice was the result of a previous challenge that the team had encountered, in which an ML model was developed using a dataset that lacked documentation about its provenance. As P15 explained, \textit{``It's very useful data, but it's useless just because there was no documentation on it.''} It seemed that the team was actively working to implement standardized data documentation approaches due to this experience. \looseness=-1

Although participants had dedicated locations for data documentation, they still resorted to finding information about datasets in documents created for other purposes such as PowerPoint presentations, team newsletters, reports, emails, scripts, and GitHub repositories. Participants recognized that data documentation was not the primary objective of these documents: \textit{``Those documents, they are not created for the dataset explicitly. They are mainly for like introducing what we have done, what we have achieved''} (P3).

While participants acknowledged that these documents were not ideal, they did not see the need to change their current approaches to data documentation. In fact, many ML scientists and data scientists felt that scripts---that is, code created to pull, preprocess, and model data---were the best form of data documentation: \textit{``The way you sort of figure out what's going on is... looking at the code that generates [the dataset] rather than referring to some specific documentation''} (P4). Another data scientist said that \textit{``the script should capture everything''} (P2), while a third believed that it was impossible to document everything of importance and that even if it were possible, \textit{``the code still remains the source of truth''} (P1). This sentiment was tied to concerns that documentation would become outdated if it were separated from code.\looseness=-1

Team B used GitHub repositories as a source of makeshift data documentation. Repository content included scripts, metadata (e.g., versioning details, logs, messages), a README file and, if small enough, datasets. Participants mentioned liking GitHub repositories because they track changes and provide access to older versions of scripts. In fact, many participants not on this team said favorable things about GitHub repositories for this reason. When asked what information would best document their dataset, a member of Team B said \textit{``the best would be a link to the GitHub repository... reveals like the attribution, the license, how it was collected and so on, so that would do a lot of the job by itself basically''} (P5).\looseness=-1

Despite the existence of some data documentation, participants said that the practice of asking others for information instead of reading the documentation was common: \textit{``most people glance at the wiki page, and then they set up a meeting with you so that they can ask you questions, because they don't bother to read the whole wiki page''} (P6). In fact, participants even described doing this themselves: \textit{``I tried to sort of get as much information as I could before he left,''} said P4 about a colleague who was retiring and had passed on an ML project to them.

When reflecting on their current approaches to data documentation, participants considered them to be incipient and informal. Despite stating that they were actively working to increase data documentation, participants on Team B described their practices along the lines of  \textit{``very, very ad hoc''} (P5) and \textit{``for now our documentation is fairly inexistent''} (P8). P11 said about the text-file datasheets maintained by their team, \textit{``it's something... but it is fairly naive.''} A couple of participants even admitted that they were unaware of how other people documented datasets, evidencing the unstandardized nature of their current approaches to data documentation. For example, P11 said, \textit{``I don't know what our researchers do for their own sake in documenting how they use datasets.''}\looseness=-1

\subsubsection{Myopic Approaches}     \label{PracticalApproach}

Participants frequently emphasized practicality and utility. For them, practicality meant that the processes by which data documentation is created should be efficient, while utility meant that the resulting documentation should contain the minimum amount of information necessary for them to determine whether they could use a dataset for a specific purpose. Participants felt that data documentation should provide short-term functional benefits that would outweigh the effort required to create it. Strikingly, they did not make the connection between the datasheets for datasets questions and their responsible AI implications.

Participants overwhelmingly prioritized efficiency, as evidenced by their enthusiasm for automated adaptation to different contexts and integration into their existing tools and workflows (discussed further in Section~\ref{Desiderata}). They felt that creating data documentation takes time away from more valued tasks: \textit{``My job is to ship stuff. Nobody rewards me for filling out paperwork. So you know, I want something I can get in and get done as quickly as possible''} (P6). Because creating data documentation was viewed as overhead, participants wanted to reduce the time and effort required to do it.\looseness=-1

Several participants said that for data documentation to be satisfactory, it would need to contain the minimum amount of information necessary for them to determine whether they could use a dataset for a specific purpose, which, for them, meant whether they had permission to use it according to internal privacy policies:  \textit{``I just would want to see what is the most relevant... like you need to know at least this before using this [dataset]''} (P15). As described by P10,  \textit{``We don't need to know... a lot of these details. Uh, good to have. Definitely not necessary to have.''} If information was not relevant for them to answer the question, ``What are approved uses of this dataset?'' then participants typically considered it less essential to document.

Participants stated that their current approaches to data documentation were enough to address their own needs as dataset creators. As one participant explained, \textit{``I just do as I need. The minimum requirement. Yeah, for my own reference''} (P3). Participants often did not look at data documentation from the perspective of dataset consumers. When asked whether the information that their team currently documented was sufficient, P15 replied, \textit{``we think it's enough for us and if there is something missing, we change it.''}\looseness=-1

At the same time, participants also noticed insufficiencies in their current approaches. One participant, who used a note-taking application and comments throughout their code to create data documentation for themself, said that their approach would be ineffective for others: \textit{``I write [notes] down, but just at the level that I know I will know... my documentation may not be serving properly for others''} (P2). Another participant expressed concerns about documenting fairness issues with datasets: \textit{``There's aspects to where I think as data scientists we fall short in terms of documenting known bias in a dataset''} (P6).

Despite highlighting the need for increased data documentation, participants felt that creating data documentation might not be worthwhile, unless the resulting documentation provided sufficient benefits. One participant explained,  \textit{``I need to make sure it's worth the effort, in terms of other people are going to use it or I myself am going to use it in the future''} (P2). If they were going to dedicate time to answering a list of questions about their dataset, the resulting documentation needed to serve a purpose. For example, when asked what would motivate them to use datasheets for datasets, P1 said,  \textit{``if I use less time to fill it in, and I feel that the information that I fill in will help me in my future workflow.''} As this participant alluded, most participants focused myopically on how data documentation might address their own needs as dataset creators, without taking into account either the benefits for other stakeholders or the responsible AI implications of the questions that they were asked to answer.\looseness=-1

\subsection{ML Practitioners' Data Documentation Perceptions} \label{RQ2}\label{OverallPerceptions}

With this background understanding of participants current approaches to data documentation, we used the full list of questions from datasheets for datasets \citep{gebru2018datasheets} to support reflection and the elicitation of participants' data documentation perceptions, needs, challenges, and desiderata. Overall, after completing the data documentation exercise, participants saw data documentation as useful, and their perceptions were largely positive. We present these perceptions next.\looseness=-1

\subsubsection{Reducing the Risk of Losing Information} Participants noted that information that is implicit or tacit is at risk of being lost if it is not documented:  \textit{``As soon as somebody leaves the team... that whole knowledge goes away and then there is no archive or there's no auditability of what they did and it's lost information''} (P14). The information covered by the datasheets for datasets questions  \textit{``would certainly reduce the need to ask questions to people... or make assumptions about poorly documented data that end up being incorrect''} (P4). Access to accurate and complete information about datasets was a powerful motivator for creating data documentation. Participants were able to recall or imagine situations in which they did not have access to the information that they needed about datasets and wished that it had been recorded somewhere for them to reference. \looseness=-1

\subsubsection{Facilitating Dataset Discovery} As highlighted by P11, when participants did take the perspective of dataset consumers, they felt that the datasheets for datasets questions would help them learn about other datasets and make informed decisions about whether specific datasets meet their needs: \textit{``I would find these [datasheets] really useful for finding a dataset that was suitable for my purpose and making sure that I really understand what it contains. Sometimes there can be a gap between what you think is in this dataset and how it was collected, and what's actually there.''}  The ability to discover datasets was closely tied to preventing repeated work. As P7 said,  \textit{``It would save time overall if data scientists knew what data is out there, be able to reuse datasets... Build on what others have done.''} A couple of participants shared this sentiment, and P14 explained that data scientists do repeat work that could be prevented if they did not lack access to the datasets that were used to develop or evaluate existing models. In other words, participants identified data documentation as a way to facilitate the collaborative sharing of datasets.

\subsubsection{Helping with Onboarding} Data documentation was also discussed as a convenient and easy way to communicate information about datasets to new team members or outside teams as part of the process of onboarding onto a dataset. Sharing this information was described as \textit{``handy''} for new team members (P8) and \textit{``helpful''} for a new project manager (P3). Part of this perceived benefit was related to the ways in which data documentation might save time that participants typically sacrificed to answering questions from new team members or outside teams. P6 summarized two purposes of the data documentation succinctly: \textit{```Keep me out of trouble and keep me from being interrupted.' If you do that, then it's useful.''}

\subsubsection{Preventing Liability}
P5 thought that data documentation might serve as a safety net, allowing dataset creators and dataset consumers to fear liability a little less:  \textit{``I just don't want us to be in a situation where actually engineers are fearful of working with data because they don't want to be hauled up in front of a disciplinary committee because they did one thing wrong, and hopefully if we have the datasheets that will actually give people more confidence.''} Additional testament to the perceived value of data documentation as a way to prevent liability was that participants asked for more of this information to be covered by the datasheets for datasets questions.

\subsubsection{Prompting Critical Thinking} Lastly, participants mentioned that completing the data documentation exercise prompted them to think critically, even beyond the datasets on which they chose to focus: \textit{``I honestly thought this was all thought provoking, and I was thinking about other datasets as I was filling this out... this [question] will apply to this one [dataset], this [question] will apply to this one [dataset]''} (P5). Another participant viewed careful evaluation of the information covered by the datasheets for datasets questions as beneficial: \textit{``I like people working with data thinking about these things, even if they don't know how to answer''} (P7).

\subsection{ML Practitioners' Data Documentation Needs and Challenges}    \label{Challenges}

By triangulating participants' answers to the datasheets for datasets questions with their subsequent reflections in the follow-up interviews, we were able to identify participants' data documentation needs and challenges. Participants had difficulties understanding the responsible AI implications of the datasheets for datasets questions, what counts as a dataset, the target audience for data documentation, and how in-depth to answer the questions. In addition, participants generally had a hard time answering some of the questions, as well as communicating and collaborating with the different stakeholders who interact with their chosen datasets at different points in time. Furthermore, differences between participants' contexts (e.g., the types of datasets with which they worked, their scopes of dataset distribution) evidenced the need for data documentation frameworks to be adaptable based on context-specific factors, which participants mentioned wanting as well.

\subsubsection{Missing the Connection to Responsible AI}    \label{RAIconnection}

Although data documentation frameworks are often motivated from the perspective of responsible AI \cite{aboutml19}, participants had difficulties making the connection between the datasheets for datasets questions and their responsible AI implications. 
Participants had to be prompted, some more than once, before they made the connection. One participant, who had undergone training so that they could become a designated team resource and contact for responsible AI, eventually mentioned it as a benefit, but only after listing other benefits and after being prompted. Some participants highlighted some of the datasheets for datasets questions as less useful, despite their direct relevance to responsible AI. Other participants, however, thought that the questions could be used to foster awareness and education about responsible AI.\looseness=-1

Some participants questioned the usefulness of those questions that were directly relevant to responsible AI. For example, multiple participants had difficulties or expressed resistance when answering the questions about dataset use. P11 felt that documenting previous uses of their dataset in response to question 5.1 \textit{``could stifle creativity... I won't perhaps be as open to thinking about using it for other purposes.''} Participants also found question 5.4 challenging to answer. This question asks, ``Is there anything about the composition of the dataset or the way it was collected and preprocessed/cleaned/labeled that might impact future uses? For example, is there anything that a future user might need to know to avoid uses that could result in unfair treatment of individuals or groups (e.g., stereotyping, quality of service issues) or other undesirable harms (e.g., financial harms, legal risks)? If so, please provide a description. Is there anything a future user could do to mitigate these undesirable harms?'' P2 said that this question was \textit{``too broad,''} P12 found it \textit{``hard to think of future uses,''} and, similarly, P10 said that they \textit{``cannot predict the future.''}  \looseness=-1

In addition, question 5.5 asks, ``Are there tasks for which the dataset should not be used? If so, please provide a description.'' Five participants answered this question with some version of ``no.'' Other participants only discussed restrictions imposed by policies or regulations. For example, P14 said, \textit{``It should not be used for any individual advertising and targeting purposes.''}\looseness=-1

Some participants also had difficulties with the questions that were related to potential impacts. Question 3.11 asks, ``Has an analysis of the potential impact of the dataset and its use on data subjects (e.g., a data protection impact analysis) been conducted? If so, please provide a description of this analysis, including the outcomes, as well as a link or other access point to any supporting documentation.''  In response to this question, P6 exclaimed, \textit{``Why am I going to be speculating on potential impact of the data on subjects of an asset that I created? I should be documenting what are approved uses of the data, because any data, you could abuse in terms of the harm you could cause.''} Meanwhile, question 2.12 asks, ``Does the dataset contain data that, if viewed directly, might be offensive, insulting, threatening, or might otherwise cause anxiety? If so, please describe why.'' P10 debated the inclusion of this question, saying,  \textit{``There's no toolbox that's going to give this to me and it's not standard to do this and I don't think it should be standard just because it's very dataset-dependent.''}\looseness=-1

Other participants saw the potential for the datasheets for datasets questions to foster awareness and education about responsible AI: \textit{``So you can ask those... leading questions to a data owner, which will make them aware about these responsible AI techniques''} (P14). One participant (P9) suggested that data documentation could even go a step further: make people aware of tasks that support responsible AI and then direct them to tools that support those tasks. The fact that so many participants missed the connection between the questions that they were asked to answer and their responsible AI implications suggests the need for data documentation frameworks to not only link to responsible AI resources and tools, but to provide more actionable guidance on how the characteristics of datasets might result in harms and how these harms might be mitigated.\looseness=-1

\subsubsection{Uncertainty about What Counts as a Dataset}     \label{DefineDataset}

Some participants were uncertain about what counts as a dataset. One reason for this lack of clarity was because their chosen datasets were produced by combining features or streams from other datasets, and it was unclear at which level of this combination process their datasets should be documented. Four participants discussed their chosen datasets as combinations of two or more other datasets. For example, P9 and P10's dataset was composed of a larger dataset containing one data type and a smaller related dataset containing a different data type. At the beginning of their follow-up interview, P9 said, \textit{``I did it [the data documentation exercise] for both, but, yeah, I think sometimes I mix them. Sometimes I only talk about one, and sometimes I talk about the other.''}\looseness=-1

It became clear that participants thought that their answers to the datasheets for datasets questions should describe the state of their chosen datasets at that point in time. Consequently, at least five participants expressed confusion about the temporal aspects of the questions:  \textit{``Sorry, I'm back to mixing my thinking. Some of them [the questions] had applicability because I was thinking of the underlying data''} (P6). This confusion may have been an artifact of our methods, specifically the fact that the data documentation exercise only afforded participants the opportunity to document their chosen datasets at a single point in time. Without clear instructions that articulated this intention, participants may have become confused.\looseness=-1

Lastly, one participant completed the data documentation exercise for multiple datasets at once. They voiced confusion over what was meant by a dataset and chose to answer the questions about a type of dataset rather than a single dataset:  \textit{``The concept of dataset wasn't very clear, whether it's like a kind of data is a dataset? Or a file is a dataset?''} (P15). Participants' uncertainties about what counts as a dataset suggests the need for data documentation frameworks to include specific guidance on how to handle streaming datasets, combined datasets, and datasets that are not self contained.\looseness=-1

\subsubsection{Uncertainty about the Target Audience for Data Documentation}   \label{Audience}
As participants completed the data documentation exercise, they mainly thought about information that would address their own needs and did not provide information that someone unfamiliar with their chosen datasets might need to know.
This was particularly pronounced when asked which of the datasheets for datasets questions were helpful or not. For example, P8 evaluated question usefulness through the lens of their team's specific desire for data provenance details: \textit{``All of these questions that say `is there?' `are there?' [Section 2] I guess they're interesting in themselves, um... I don't think it's really useful for the purpose of provenance, which is the question I'm interested in''} (P8). \looseness=-1

Participants used phrases like \textit{``may not be very meaningful for us''} (P2), \textit{``not that relevant to me''} (P8), and \textit{``It doesn't matter exactly... I don't need to know''} (P11) when giving feedback on the datasheet for datasets questions. Participants did recognize that those questions that did not apply to their chosen datasets might be applicable to others' datasets (e.g., \textit{``I thought all of them were useful to some datasets, it's just not mine''} [P5]). The fact that the information covered by some of the questions would be useful for dataset consumers rarely came up unless participants were discussing what information might be important to know about their chosen datasets.\looseness=-1

Participants' answers also hinted at their difficulties empathizing with people unfamiliar with their chosen datasets. For example, many participants failed to describe, define, or explain terms, data content, or tools in their answers to the datasheets for datasets questions. These failures to provide detailed answers could indicate that participants had difficulties taking the perspective of dataset consumers. A couple of participants considered data documentation for the purpose of communicating and collaborating with dataset consumers to be novel. One participant proclaimed, \textit{``We never thought to document [the dataset] in a way that would enable those outside of [team name] to really understand without needing to ask us''} (P11). Another participant thought that sharing this information would be atypical: \textit{``Unless the datasheet travels with the data when we share with another team... but I think that's really a corner case here''} (P8).

A few participants did recognize that data documentation needed to serve as a resource for dataset consumers, in addition to addressing their own needs. P15 wondered whether they should describe terms that their team was familiar with, but that someone else might not understand, at the top in a glossary. Another participant said that the purpose of data documentation was \textit{``to follow my data resource so that other people... have all the information that they need to use it properly''} (P6). Despite participants' sporadic recognition of data documentation as a collaboration artifact, their answers to the datasheets for datasets questions suggest that they did not have this in mind when documenting their chosen datasets. As a result, data documentation frameworks could be improved by clearly specifying that the target audience should include dataset consumers, who may be unfamiliar with the datasets being documented.\looseness=-1

\subsubsection{Unclear Understanding of How In-Depth to Answer}  \label{ResponseDepth}

Participants questioned how in-depth their answers to the datasheets for datasets questions needed to be:  \textit{``At which level of granularity am I trying to fill this up?''} (P8) and  \textit{``Am I going to too much details?''} (P11). This was especially true if the information covered by a question was not tied to their direct responsibilities. P15 explained that their team creates some datasets, but simply uses other datasets. As a result, they said \textit{``I was confused to how detailed I should be in terms of the dataset for which we are users.''}

In alignment with their myopic approaches to data documentation, many participants chose not to provide detailed information in their answers. Although these participants did not experience confusion about how they should answer the questions, they assumed that succinct answers were sufficient. \textit{``Pretty easy to just say like one liner and they'll be done... I think I can get done with this datasheet by filling minimal amount of information and still be okay''} (P14). Most participants tended to only answer the first parts of the datasheets for datasets questions and not the follow-up questions directly afterward (e.g., ``describe,'' ``provide further details''). As one participant stated, \textit{``Most of [the questions] I didn't need to read the small text, the [first part] was enough''} (P6).

The following quotes are examples of participants' succinct answers to questions that asked for further elaboration. For question 1.1 (``For what purpose was the dataset created? Was there a specific task in mind? Was there a specific gap that needed to be filled? Please provide a description.''), P3 answered,
\textit{``for [dataset name] model training purpose.''} In response to question 2.5 (``Is there a label or target associated with each instance? If so, please provide a description.''), P9 responded with a generic \textit{``Yes the comments are labeled.''} Lastly, like many other participants, P14 responded with just \textit{``Yes''} to question 7.6 (``Will older versions of the dataset continue to be supported/hosted/maintained? If so, please describe how. If not, please describe how its obsolescence will be communicated to users.'')\looseness=-1

Of the 11 ``datasheets'' that we included in our data analysis, only one consistently contained detailed answers. This participant (P11) spent five to six hours completing the data documentation exercise, but attributed only one or two of those hours to answering the datasheets for datasets questions versus specifying where the information covered by the questions is currently documented. Other participants said that the exercise took them between 30 minutes to three hours. We note that participants' lack of detailed answers, while not expected or ideal, is a meaningful finding in and of itself. Importantly, it reveals that---at least when data documentation is not mandated and sufficient time is not readily available to create it---data documentation may be of low quality. This is important to keep in mind given that previous research on the use of data documentation~\cite[e.g.,][]{B21} has assumed the existence of high-quality documentation.
That said, since our goal was to use the datasheets for datasets questions only as an exercise to support reflection and the elicitation of participants' data documentation perceptions, needs, challenges, and desiderata, not for the purpose of analyzing the specific answers provided by participants, we believe that the quality of participants' answers had minimal impact on our findings. Overall, the lack of detailed answers does suggest the need for data documentation frameworks to explicitly state in their instructions the depth expected.\looseness=-1

\subsubsection{Lack of Investigation}  \label{QuestionBarriers}

When participants did not know how to answer the questions, they often left their answers blank or wrote, ``I'm unsure.'' P9 explained their approach:  \textit{``I just put for some of the questions that, `I haven't heard about that term,' or, `I haven't done that task,' but I didn't take the time to investigate what was that.''} Other participants acted similarly. When they did not know the information that they needed, they did not necessarily search to find it:  \textit{``I wasn't sure... whether I should go and find resources and find out the answers for those, or what is my responsibility?''} (P15).\looseness=-1

For the most part, when participants did search for information about their chosen datasets, they looked through existing documents, but did not seek assistance from others. In fact, only P12 sought help to confirm that one of their answers was accurate. Due to their independent approaches to documentation, participants' unknown answers remained as such, but were at least honest representations of their knowledge:  \textit{``I don't know the details, because it's like PMs, they contact some person to do that... But what I can do is... answer as of current state what I know''} (P3). Participants' tendencies not to seek assistance from others suggest that data documentation frameworks need to provide guidance on and set expectations for sharing information.\looseness=-1

\subsubsection{Lack of Collaboration}

Participants' roles tended to correlate with different stages of the dataset lifecycle. For the data documentation exercise, ML scientists and data scientists typically chose to focus on ``downstream'' datasets, far removed from dataset creation processes and ready to be used to develop or evaluate  ML models. In contrast, software engineers typically chose to focus on ``upstream'' datasets, closer to dataset creation processes and typically less processed. As a result, different participants struggled with different questions. Sometimes participants were unsure about who might know the information that they needed to answer the questions or whether they were responsible for documenting that information. Some of these challenges were explicitly referenced by participants, while others were deduced from participants' answers and their follow-up interviews. However, they all suggest the need for better communication and collaboration between the different stakeholders who interact with datasets at different points in time.\looseness=-1

Many times, participants who chose to focus on ``downstream'' datasets lacked the information that they needed to answer the questions about data collection such as questions covering how data was obtained from participants, when it was obtained, by whom, and so on. Participants tolerated these gaps in their knowledge and expressed that, to an extent, using such datasets necessitated placing faith in those who had done the ``upstream'' work. Two participants who chose to focus on datasets scraped from the web shared, \textit{``We know only this much, and we rely on the team that provides us the data, that they are going through the right process''} (P15) and \textit{``you place a degree of trust... that they've done the right thing and have not actually violated anything''} (P5). Additionally, for streaming datasets, different stakeholders are often responsible for creating and maintaining the ``upstream'' code that pulls in the data. As one ML scientist stated when they lacked information about data collection, \textit{``the people who are in charge of handling data like way before us... they are like the better people to seek the answer''} (P2).

Strikingly, participants who chose to focus on ``downstream'' datasets had a hard time understanding why someone else might want to use their chosen datasets. Because these datasets were tailored for very specific purposes, participants were \textit{``a little confused because... they [people wanting to use the data] probably will generate the data from the raw telemetry, not on top of this cooked dataset''} (P2). P1 mentioned that data scientists only use a dataset a \textit{``few times and then you move onto the next project... and then the old dataset becomes irrelevant.''}

In contrast, participants who chose to focus on ``upstream'' datasets that were used for many different purposes had difficulties answering the questions about preprocessing and labeling. In addition, a couple of participants also felt that two other questions did not apply to their chosen datasets. These were question 2.5 (``Is there a label or target associated with each instance? If so, please provide a description'') and 2.8 (``Are there recommended data splits [e.g., training, development/validation, testing]? If so, please provide a description of these splits, explaining the rationale behind them.'').\looseness=-1

Both ML scientists and data scientists, who chose to focus on ``downstream'' datasets, and software engineers, who chose to focus on ``upstream'' datasets, wanted the datasheets for datasets questions to cover information spanning the entire dataset lifecycle and, specifically, those parts of the lifecycle with which they were less familiar. As one software engineer explained, \textit{``There are upstream dependencies and then this [data] string could be used for other downstream dependencies that the datasheet does not have a lot of questions around''} (P12). In fact, there were questions on stages of the dataset lifecycle covering ``downstream'' dependencies. This participant's feedback illustrates a challenge inherent to documenting streaming datasets and the subsequent need for more communication and collaboration between the different stakeholders who interact with datasets at different points in time. \looseness=-1

Participants tended to answer the datasheets for datasets questions with information that was only relevant to the ways in which they worked with their chosen datasets. For example, when asked who created their chosen dataset, one data scientist answered, \textit{``I created this specific dataset. It's joined together from several other datasets''} (P4). At a very different stage of the dataset lifecycle, a software engineer questioned whether their chosen dataset was even an ML dataset: \textit{``I realized later on this dataset is mainly used for analytics... and not so much directly for... Machine learning models... lot of the features that end up generated for a ML models derived from these dataset but this is more like upstream in the layer.''} These findings suggest the need for data documentation frameworks to include clear instructions that emphasize the collaborative and longitudinal nature of the processes by which data documentation is created.\looseness=-1

\subsubsection{Adaptation to Different Contexts}  \label{DataDistribution}

Differences in participants' contexts---specifically, the types of datasets with which they worked and their scopes of dataset distribution---made them feel that some questions were irrelevant, inapplicable, or confusing, causing them to express the need for data documentation frameworks to be adaptable based on context-specific factors.

Participants described the distinction between internal and external distribution as an important factor that would affect data documentation. Internal distribution was viewed as distributing a dataset within the company, while external distribution was viewed as distributing a dataset outside of the company. From a privacy standpoint, both types of distribution are subject to review and approval according to internal company policies. Many participants said that the questions about distribution plans needed to be asked early on so that irrelevant or inapplicable questions could then be filtered out. As one participant described, \textit{``In an ideal world, maybe there would be a general version of this that is more targeted towards internal versus external like datasets and... you could probably lose some of the questions''} (P4). Another participant said that some of questions \textit{``might be too heavy for an internal dataset''} but that for an externally distributed dataset those questions would need to be answered \textit{``as extensively as possible''} (P14).

All but two participants did not have plans to externally distribute their chosen datasets and said that the questions about distribution plans were irrelevant or inapplicable. These participants typically felt the same way about the questions about maintenance. In contrast, the two participants who were planning to distribute their chosen dataset externally, liked the questions about maintenance. In particular, they felt that question 7.7 (``If others want to extend/augment/build on/contribute to the dataset, is there a mechanism for them to do so? If so, please provide a description. Will these contributions be validated/verified? If so, please describe how. If not, why not? Is there a process for communicating/distributing these contributions to other users? If so, please provide a description.'') was useful. These differences in participants' sentiments suggest that the perceived value of the datasheets for datasets questions is context dependent.

Different participants chose to focus on different types of datasets, with some participants choosing to focus on streaming datasets and others choosing to focus on static datasets. Because datasheets for datasets were designed for static datasets, many of the questions were challenging to answer for participants whose chose to focus on streaming datasets. Participants were unsure how to handle questions whose answers might vary over time or questions that seemed irrelevant or inapplicable to streaming datasets. For example, in reference to question 3.5, P15 asked,  \textit{```Over what timeframe was the data collected?' So, if the data collection is an ongoing process, how do we answer it? Because we started collecting the data when the company was founded.''}\looseness=-1

Due to their continually evolving nature, streaming datasets led to insufficiently detailed answers to some of the questions about dataset composition. When answering question 2.2 (``How many instances are there in total [of each type, if appropriate]?''), for example, participants often gave generic answers: \textit{``The data comes from the production traffic, the volume depends on the feature traffic and it varies over time''} (P2) and \textit{`` > 2 billion''} (P12). Overall, and unsurprisingly, participants who chose to focus on streaming datasets had a much harder time answering the datasheets for datasets questions.\looseness=-1

These findings suggest that there is no one-size-fits-all approach to data documentation. As a result, participants discussed the need for data documentation frameworks to be adaptable based on context-specific factors. While they tied this need to the fact that they felt that some of the datasheets for datasets questions did not apply to their chosen datasets, they also frequently mentioned that the full list of datasheets for datasets questions took too long to answer. P1 said,  \textit{``Literally, the number [of questions] is just mind-boggling.''} Adapting data documentation frameworks to different contexts also has the benefit of reducing the number of questions that dataset creators need to answer.\looseness=-1

\subsection{ML Practitioners' Data Documentation Desiderata}    \label{Desiderata}

Participants shared their desiderata for data documentation frameworks during the follow-up interviews, highlighting that they wanted data documentation frameworks to have interactive capabilities, integration into their existing tools and workflows, and automation to keep data documentation up to date.\looseness=-1

\subsubsection{Interactive Capabilities}   \label{Interactivity}

Interactive capabilities were one of participants' most common desires for data documentation frameworks. Several participants wanted data documentation frameworks to filter the questions that dataset creators would see based on their answers to previous questions: \textit{``ask a couple of questions and then depending on that, then you answer like a subset of questions''} (P10). P5 articulated a similar request: \textit{``That would be really nice if it [the datasheet] was just more constrained, and then you wouldn't have to answer `No' so much... just a bit of interactivity on the form. So, if you, in the combo box select this thing, then the following section adjusts to something else.''} Part of participants' enthusiasm for filtering questions might have been be due to the fact that we used the full list of datasheets for datasets questions rather than tailoring the questions. As a result, participants typically encountered several questions that did not apply to their chosen datasets or to which they answered ``no.''  \looseness=-1

A couple of participants talked at length about restricting answer choices for some of the questions. They discussed using drop-down menus and yes/no checkboxes to \textit{``shrink down the size of the document so it doesn't overwhelm somebody''} (P14). P14 went on to discuss the benefits of restricting answer choices: \textit{``The standardization of the response will help you. If you don't sanitize there, you will never be able to aggregate and give those details to somebody, so templatization would be great... like as many questions that can use drop-downs, I think you should use there, specially for yes or no.''}\looseness=-1

In addition, many participants saw search capabilities as essential for data documentation frameworks. As one participant bluntly stated, \textit{``The concept of searchability means that the information is not lost... information that is not findable does--might as well not exist''} (P1). Participants' desires for search capabilities were closely tied to their desires for tracking metadata and for automation. This was especially true for participants who saw facilitating dataset discovery as an important benefit of data documentation such as P11 who explained, \textit{``It is really important that it [the datasheet] be as machine readable and structured and automatable as possible so that we can really search and perform meta-analysis on this.''}\looseness=-1

\subsubsection{Integration into Existing Tools and Workflows}    \label{Integration}

Participants felt that integration of data documentation frameworks into their existing tools and workflows was extremely important. They wanted the processes by which data documentation is created to be integrated with the databases, cloud platforms, and analysis tools that they use. Integration was described through the lens of incentives and deterrents: \textit{``if it's somehow integrated into sort of my day-to-day workflows... was in the tools that I was already using... that wouldn't be all that onerous, but if I had to like go somewhere else and remember to do it and keep it up to date, that would be sort of a higher barrier to entry''} (P4) and \textit{``I'm more excited when we have something that's semi-automated built into the [platform name] pipelines... that's integrated into where and how I do my normal development work where it fits. Otherwise, it feels like a bolt-on''} (P6). Participants also suggested that information from data documentation could be exported: \textit{``If I fill in a field that the field can be easily transferred to some kind of database''} (P1). Another participant mentioned the possibility of developing data documentation tools that could be integrated with existing tools for developing and evaluating ML models.\looseness=-1

Some participants stressed that they wanted central locations for data documentation, where multiple people could access and add information.  One participant said, \textit{``as long as it's one... centralized place for different projects, and some other folks, they may ask, and we can send the link''} (P3). A couple of participants also felt that the datasheets for datasets questions facilitated the (much needed) standardization of data documentation, which made information easy to find: \textit{``I'm really glad we're coming up with a standard... that can be used across [company name]... it means that if I'm looking at the dataset produced by another team, I'll know exactly where to look in this datasheet to get the information I want''} (P11).  \looseness=-1

Many participants requested links to scripts and other resources. ML scientists and data scientists were excited about the potential for data documentation to connect ``upstream'' details about datasets with the ML models that use those datasets ``downstream.'' P2 said, \textit{``I think good information would be like maybe what are the raw streams you used and also... the location of the stream and also the script that's generated the stream.''} Meanwhile, P3 said that they wanted to know \textit{``which models uses this dataset, and if there's any link document... What are the like accuracy numbers from those previous models?''} They went on to mention that the datasheets for datasets questions did not ask about the ML models that use the dataset being documented and suggested adding questions on model statistics.\looseness=-1

Participants also wanted to be able to track metadata, mentioning the need to track versions of their datasets and access changelogs. One participant shared, \textit{``I really care about the history, because for this dataset, we did like I don't know how many rounds of model refresh, model structure update''} (P3). Overall, participants liked the metadata that GitHub gathers as a version control platform and one participant even went as far as to say, \textit{``I would love to have some sort of Git-like repository where you submit raw data, and you have all this information about it, and if someone cleaned it, that version, information on how it was cleaned, and the script. And if someone does this again, repeat''} (P1). Another participant felt strongly that dataset creators should know who the consumers of their datasets are. They suggested that metadata might even provide dataset creators with an incentive to document their datasets: \textit{``On top of... these datasheets, it is essentially metadata for the data and... that metadata then becomes searchable or question answerable... provide a capability, then people will be hooked onto it [the datasheet] because they would love to give more data.''} (P14).

\subsubsection{Automation}    \label{Automation}
Many participants wanted data documentation frameworks to be automated wherever possible, so as to reduce the effort required to create data documentation and to keep data documentation up to date. The ability to automatically answer the datasheets for datasets questions played an important role in some participants' perceptions of the questions' value: \textit{``We would say `can we populate easily these things?' and if we can, our threshold for keeping the question would be lower''} (P8). It also impacted their perceptions of the benefits of data documentation: \textit{``automated, and if it's not, as the friction [of filling it out] goes down, the amount of support goes up. As the friction [of filling it out] is higher, the amount of positive feedback I have to get from producing the asset needs to be higher''} (P6). Automation was therefore closely tied to participants' emphasis on practicality and utility, as described in Section~\ref{PracticalApproach}.

A few participants saw automation as way to keep data documentation up to date. The fear that some information might become outdated was prevalent, especially for participants who chose to focus on streaming datasets: \textit{``There's a risk of trying to document it here versus reference where the most recent should be stored''} (P6). Such concerns prompted participants to question how data documentation is maintained and whether the information documented could be trusted: \textit{```How do the answers in here sort of stay up to date?' If I'm reading one of these that was created like two years ago, I wouldn't have like the greatest confidence that the content in there is still current.''} (P4). P1 reasoned that data documentation would be reliable if it was created automatically, saying that \textit{``because it could be automated, it would likely be more accurate.''}

\section{Discussion}
\label{Discussion}

Our findings provide rich grounds for improving data documentation frameworks. In particular, data documentation frameworks would benefit from several modifications that emerged when answering our second research question. One of our most concerning findings is that participants did not make the connection between the datasheets for datasets questions and their responsible AI implications, as discussed further below. Following that discussion, we then present seven design requirements for future data documentation frameworks and discuss the limitations of our study.

\subsection{Data Documentation and Responsible AI}

Data documentation frameworks are often motivated from the perspective of responsible AI. Data documentation is viewed as a way to increase transparency, mitigate harms, enable a more nuanced understanding of the extent to which ML models might generalize, and prompt reflection on the consequences of the choices made throughout the dataset lifecycle~\cite{gebru2018datasheets,holland2018dataset,bender2018data,aboutml19}. However, participants in our study often failed to make the connection between the questions that they were asked to answer and their responsible AI implications.  Rather, as discussed in Section~\ref{PracticalApproach}, participants prioritized information about whether a dataset was approved for use. In other words, participants were more focused on documenting information that would allow them to determine whether they or others \emph{could} use a dataset for a specific purpose as opposed to whether they or others \emph{should} use it and what might go wrong if they were to do so.  This is worrisome in light of the frequency with which problematic or inappropriate datasets lead to performance disparities between demographic groups, models that don't generalize, and other common ML failures~\cite{PRB+20}. The inability to recognize when a dataset is problematic or inappropriate, meaning that it should not be used, mirrors a broader problem within technology industry of differentiating between technologies that \emph{can} be built versus technologies that \emph{should} be built~\cite{BS11,S19,BB+20}.

This tension between \emph{could} and \emph{should} played out in participants' myopic approaches to data documentation, and in the (sometimes contradictory) views that they expressed about whether their current approaches were sufficient. As discussed in Section~\ref{PracticalApproach}, participants tended to document only enough information to address their own needs as dataset creators, even while acknowledging that this approach would likely not satisfy the needs of dataset consumers. While some participants noted that some of the datasheets for datasets questions might prompt critical thinking or cover information that would be useful for dataset consumers, deliberate reflection on datasets and transparency around the processes by which they are created were not primary motivators, nor were the needs of others. \looseness=-1

The benefits that participants might gain from more deliberate reflection were also at odds with their strong desires for automation (see Section~\ref{Automation}). While some forms of automation may be possible without compromising reflection---for example, automatically extracting information from privacy reviews or other parts of existing workflows---\citet{gebru2018datasheets} specifically designed many of the datasheets for datasets questions to cover information that cannot be automatically extracted from a dataset such as whether there is ``anything about the composition of the dataset or the way it was collected and preprocessed/cleaned/labeled that might impact future uses'' or whether there are ``tasks the dataset should not be used for.''

Participants also expressed resistance when answering the questions about dataset use, saying that they \textit{``cannot predict the future''} (P10), as discussed in Section~\ref{RAIconnection}. This mirrors the current debate within the ML research community over the extent to which researchers should speculate about the broader impacts of their research. As pointed out by \citet{BOC20}, it is difficult to foresee failures and harms that might arise in new contexts, and attempts to do so are often limited by ``failures of imagination.'' How best to train ML researchers and practitioners to engage in creative speculation~\cite{F21} or to otherwise anticipate potential consequences of their work is an area where more research is needed.\looseness=-1

\subsection{Design Requirements}
\label{DesignRequirements}

Taken together, our findings suggest seven design requirements for future data documentation frameworks, which we describe below. \highlight{All of these design requirements relate to the processes by which data documentation is created, with a focus on helping dataset creators produce high-quality data documentation that can act as a mode of communication and collaboration between themselves and dataset consumers. In addition, design requirements 5--7 are also about the ways that data documentation frameworks might support communication and collaboration during the processes by which data documentation is created and, ultimately, with dataset consumers.} \\

\noindent{\textbf{\highlight{1.} Make explicit the connection between data documentation and responsible AI.}} Although participants did not make the connection between the datasheets for datasets questions and their responsible AI implications, they were often able to do so after being prompted. This suggests that data documentation frameworks should make explicit the connection between the questions that datasets creators are asked to answer and their responsible AI implications. Dataset creators should be provided with more actionable guidance on how the characteristics of datasets might result in harms and how those harms might be mitigated. For example, if dataset creators answer ``yes'' to the question, ``Does the dataset identify any demographic groups, groups that are protected by law, or groups that may be considered sensitive (e.g., age, gender, disability status)?'' they could be prompted to consider the likelihood that the distribution of these groups will cause performance disparities and presented with resources on how to evaluate ML models for such disparities~\cite{BG+21}. Going a step further, if a dataset identifies demographic groups based on race, its creators could be prompted to reflect on the reductiveness of racial categories and asked to make explicit the assumptions behind the racial categories reflected in the dataset~\cite{miceli2021documenting}.  As another example, depending on the answer given to ``Who was involved in the data collection process and how were they compensated?'' dataset creators could be pointed toward resources on how to set fair wages for crowdworkers~\cite{ST+18} or, following \citet{miceli2021documenting}, prompted to interrogate the asymmetrical power relationships between those who request or commission datasets and those who label them, as well as the implications of those relationships. \highlight{By helping dataset creators make the connection between data documentation and responsible AI, data documentation frameworks can encourage more deliberate reflection on datasets and on the consequences of the choices made throughout the dataset lifecycle.\looseness=-1}\\

\noindent{\textbf{\highlight{2.} Make data documentation frameworks practical.}} Participants' current approaches to data documentation highlight the need to emphasize practicality and utility in future data documentation frameworks. Participants described having many demands on their time, and were candid about the likelihood that they would create data documentation. They recognized several benefits of data documentation, as discussed in Section~\ref{OverallPerceptions}, but, at the same time, said that if the effort taken to create data documentation outweighed those benefits,
  they would be less motivated to do so. However, if creating data documentation could speed up approval processes and if approvals were appended to data documentation, then they would be more likely to invest the time.\\

  \noindent{\textbf{3. Adapt data documentation frameworks to different contexts.}} \citet{gebru2018datasheets} note that the datasheets for datasets questions are not meant to be prescriptive and should be adapted to different domains, organizational infrastructures, and workflows. Indeed, as discussed in Section~\ref{Integration}, participants expressed the need for adaptation to these contexts, as well as in Section~\ref{DataDistribution}, the need for data documentation frameworks to be adaptable based on other context-specific factors such as whether a dataset is static or streaming or how broadly a dataset will be distributed. Depending on their contexts, different participants perceived the value of the datasheets for datasets questions differently. Our findings therefore suggest the need for data documentation frameworks to filter the questions that dataset creators would see based on their answers to previous questions. Adapting data documentation frameworks to different contexts has the additional benefit of reducing the number of questions that dataset creators need to answer, as well as potentially reducing confusion and making sure that the questions are maximally beneficial for both dataset creators and dataset consumers. In addition, specific guidance on how to handle streaming datasets, combined datasets, and datasets that are not self contained might help dataset creators better understand what counts as a dataset.\looseness=-1\\

\noindent{\textbf{4. Don't automate away responsibility, but do support simple tasks with automation.}} While participants were excited about automation, it is important to keep in mind that automation can defeat one of the main goals of data documentation---to encourage responsible AI practice through more deliberate reflection on datasets. However, some amount of automation could help keep data documentation up to date, particularly when documenting streaming datasets, or at least signal when updates need to be made. Similarly, automation could be used to extract some types of metadata or to append privacy or legal approvals to data documentation.\looseness=-1\\

\noindent{\textbf{5. Clarify the target audience for data documentation.}} Participants expressed uncertainty about the target audience for data documentation. They had difficulties prioritizing the needs of dataset consumers and thinking of future uses for their chosen datasets. Some participants mentioned that data documentation could help with onboarding onto a dataset, indicating they were thinking about the needs of others. However, other participants focused on their own needs as dataset creators. Data documentation frameworks should clarify the target audience for data documentation, including dataset consumers. This is particularly important as participants' answers to the datasheets for datasets questions often lacked information that someone unfamiliar with their chosen datasets might need to know.\\

\noindent{\textbf{6. Standardize and centralize data documentation.}} Participants suggested standardizing data documentation as a way to make information easier to find. They also suggested having central locations for data documentation---perhaps even searchable, central repositories---where multiple people could access and add information, facilitating dataset discovery and fostering communication and collaboration between dataset creators and dataset consumers. This would increase the efficacy of data documentation as a boundary object~\cite{lee2007boundary}.\\

\noindent{\textbf{7. Integrate data documentation frameworks into existing tools and workflows.}}
Participants pointed out the importance of integrating data documentation frameworks into their existing tools and workflows. While this finding generally aligns with previous research~\cite{amershi_software_2019,hohman_understanding_2020,patel_gestalt_2010,madaio2020co}, our findings point to specifics of how this integration might happen and what kinds of information could be brought together. Many participants wanted links to scripts and other resources, with some participants envisioning data documentation serving as way to connect ``upstream'' details about datasets with the ML models that use those datasets ``downstream.'' Built-in reminders about when in the dataset lifecycle specific questions should be answered may also be useful. While integration would likely benefit individual ML practitioners, it might also prompt dataset creators to seek assistance from others when creating data documentation rather than leaving answers blank. This would reduce the burden on dataset creators, as well as resulting in higher-quality documentation.\looseness=-1

\subsection{Limitations}
\label{Limitations}

Our study was designed to understand whether data documentation frameworks meet the needs of ML practitioners. Specifically, we set out to understand ML practitioners' data documentation perceptions, needs, challenges, and desiderata, with the ultimate goal of deriving design requirements that can inform future data documentation frameworks. To do this, we gave participants the full list of questions from datasheets for datasets~\cite{gebru2018datasheets} and asked them to answer the questions to the best of their abilities with respect to a dataset with which they had worked recently. Because of the artificial manner in which the datasheets for datasets questions were presented to participants, our study should not be viewed as an evaluation of datasheets for datasets. While \citet{gebru2018datasheets} note that the list of questions is not meant to be prescriptive and should be adapted to different contexts, we provided the full list to participants. We also omitted guidance on when each question should be answered and only afforded participants the opportunity to document their chosen datasets at a single point in time. We also provided the questions in a Word document rather than integrating them into participants' existing tools and workflows. As a result, participants' experiences in our study are likely not reflective of their experiences in practice, meaning that although our study contributes insights that should inform future data documentation frameworks, additional research is needed to understand how data documentation frameworks are used in practice.

Because participants were asked to answer the datasheets for datasets questions as part of a study rather than as part of their day-to-day work, some participants allocated less time to the exercise than they might have otherwise. Participants were also unlikely to seek assistance from others when they did not know how to answer the questions. This finding confirms that the processes by which data documentation is created are collaborative and that data documentation can serve as a boundary object~\cite{star1989institutional}. At the same time, this finding explains why some participants' answers were short, insufficiently detailed, or left blank. That said, since our goal was to use the datasheets for datasets questions only as an exercise to support reflection and the elicitation of participants' data documentation perceptions, needs, challenges, and desiderata, not for the purpose of analyzing the specific answers provided by participants, we believe that the quality of participants' answers had minimal impact on our findings.\looseness=-1

Datasheets for datasets were designed for static datasets, and one of our sampling criteria was that participants had to work with static datasets. However, multiple participants ended up choosing to focus on streaming datasets. While these participants struggled with some of the datasheets for datasets questions, including them in our study enabled us to gain a deeper understanding of the needs of practitioners who work with streaming datasets.

Finally, all participants were from a single large, international technology company. While we believe that most of our findings hold more generally, it is possible that some might reflect the specific culture and practices at this company.

\section{Conclusion}

We undertook a qualitative study with 14 ML practitioners to understand their data documentation perceptions, needs, challenges, and desiderata. By conducting semi-structured interviews and asking them to answer a list of questions taken from datasheets for datasets~\cite{gebru2018datasheets}, we found that their current approaches to data documentation are largely ad hoc and myopic in nature, with participants prioritizing information about whether they or others \emph{could} use a dataset for a specific purpose over a deeper consideration of whether they or others \emph{should} use it and what might go wrong if were to do so. Relatedly, but surprisingly, we also found that few of our participants made the connection between the questions that they were asked to answer and their responsible AI implications. We therefore recommend that future data documentation frameworks provide more actionable guidance on how the characteristics of datasets might result in harms and how these harms might be mitigated. Participants also struggled with some of the questions. This was because they had difficulties understanding how in-depth to answer the questions, did not seek assistance from others when they did not know how to answer the questions, and, in some cases, were uncertain about what counts as a dataset. Participants were frustrated with the length of the data documentation exercise and wondered how data documentation might be kept up to date. They therefore expressed needs for data documentation frameworks to be adaptable to
their contexts, integrated into their existing tools and workflows, and automated wherever possible. Based on our findings, we derived seven design requirements for future data documentation frameworks.

Moving forward, we see several avenues for future work. First, there is a clear need to bridge the gap between the ideal of data documentation as way to encourage responsible AI practice and practitioners' perceptions of data documentation. Second, our findings suggest that there is no one-size-fits-all approach to data documentation. As a result, data documentation frameworks need to be adaptable---perhaps automatically---based on context-specific factors. In particular, given that so many participants chose to focus on streaming datasets despite our sampling criteria, there may be a need for data documentation frameworks that are specifically intended for documenting streaming datasets. Finally, although we used the datasheets for datasets questions as an exercise to support reflection and the elicitation of participants' data documentation perceptions, needs, challenges, and desiderata, our study should not be viewed as an evaluation of datasheets for datasets. Understanding how data documentation frameworks are used in practice remains an important direction for future work.\looseness=-1

\section*{Acknowledgements}

We thank all of the participants in our study.  We are grateful to the CSCW reviewers who provided constructive feedback that led to a partial reorganization of our results. Thanks to Emily Bender, Sarah Bird, Kate Crawford, Hal Daum\'{e} III, Timnit Gebru, Mike Hind, Josh Hinds, Eric Horvitz, Martha Laguna, Michael Madaio, Meg Mitchell, Jamie Morgenstern, Ben Noah, Mehrnoosh Sameki, Kush Varshney, Briana Vecchione, Microsoft's Aether Transparency Working Group and Aether Fairness \& Inclusiveness Working Group, the Data Nutrition Project team, and the Partnership on AI ABOUT ML team for many valuable conversations over the years.  Special thanks to Yang Trista Cao and Hal Daum{\'e} III for allowing us to use the Gender Inclusive Coreference datasheet as an example and to Lisa Egede and Hari Subramonyam for providing feedback on interview protocols.  Much of this work was conducted while Liz B. Marquis was an intern at Microsoft Research.

%%
%% The next two lines define the bibliography style to be used, and
%% the bibliography file.
%%% -*-BibTeX-*-
%%% Do NOT edit. File created by BibTeX with style
%%% ACM-Reference-Format-Journals [18-Jan-2012].

\bibliographystyle{ACM-Reference-Format}
%\bibliography{sample-base}

%%
%% If your work has an appendix, this is the place to put it.
\appendix \label{Appendix}

\end{document}